\documentclass[aps,prl,twocolumn,superscriptaddress,showpacs,floatfix]{revtex4}
\usepackage{graphicx}
\usepackage{dcolumn}
\usepackage{bm}
\begin{document}
\title{
        Interplay between structure and magnetism in
        Mo$_{12}$S$_{9}$I$_{9}$ nanowires }

\author{Teng Yang}
\affiliation{Physics and Astronomy Department,
             Michigan State University,
             East Lansing, Michigan 48824-2320 }

\author{Shinya Okano}
\affiliation{Physics and Astronomy Department,
             Michigan State University,
             East Lansing, Michigan 48824-2320 }

\author{Savas Berber}
\affiliation{Institute of Physics, University of Tsukuba,
             1-1-1 Tennodai, Tsukuba, Ibaraki 305-8571, Japan }

\author{David Tom\'anek}
\email[E-mail: ]{tomanek@msu.edu}
\affiliation{Physics and Astronomy Department,
             Michigan State University,
             East Lansing, Michigan 48824-2320 }

\date{\today}

\begin{abstract}
We investigate the equilibrium geometry and electronic structure
of Mo$_{12}$S$_{9}$I$_{9}$ nanowires using {\em ab initio} Density
Functional calculations. The skeleton of these unusually stable
nanowires consists of rigid, functionalized Mo octahedra,
connected by flexible, bi-stable sulphur bridges. This structural
flexibility translates into a capability to stretch up to
${\approx}20$\% at almost no energy cost. The nanowires change
from conductors to narrow-gap magnetic semiconductors in one of
their structural isomers.
\end{abstract}




\maketitle



One of the major challenges in the emerging field of molecular
electronics is to identify conducting nanostructures with desired
electronic properties, which are stable and easy to manipulate.
Due to their high stability and favorable electronic
properties~\cite{RSaito98}, carbon nanotubes~\cite{Iijima91} have
been discussed extensively as promising candidates for molecular
electronics applications. A major drawback is the current
inability to synthesize or to isolate nanotubes with a given
diameter and chiral angle, which determine their metallic or
semiconducting nature. Moreover, single-wall carbon nanotubes are
hard to isolate from stable bundles, which form spontaneously
during synthesis \cite{Thess96}. In this respect, recently
synthesized nanowires based on Mo chalcogenides, such as
Mo$_{12}$S$_{9}$I$_{9}$, seem to offer several advantages by
combining uniform metallic behavior, atomic-scale perfection and
easy dispersability \cite{{VNicolosi05cpl},{VNicolosi05jpcb}}.

Using {\em ab initio} density functional calculations, we study
the suitability of Mo$_{12}$S$_{9}$I$_{9}$ nanowires as potential
building blocks of electronic nanocircuits. In terms of binding
energy per atom, we find these nanowires to be almost as stable as
carbon nanotubes. The nanowire skeleton consists of rigid,
functionalized Mo octahedra,
connected by flexible, bi-stable sulphur bridges. We find this
structural flexibility to translate into an unusual capability to
stretch up to ${\approx}20$\% at virtually no energy cost, as a
nanostructured counterpart of an accordion. Our calculations
suggest that the nanowires change from conductors to narrow-gap
magnetic semiconductors in one of their structural isomers.

\begin{figure}[b]
\includegraphics[width=1.0\columnwidth]{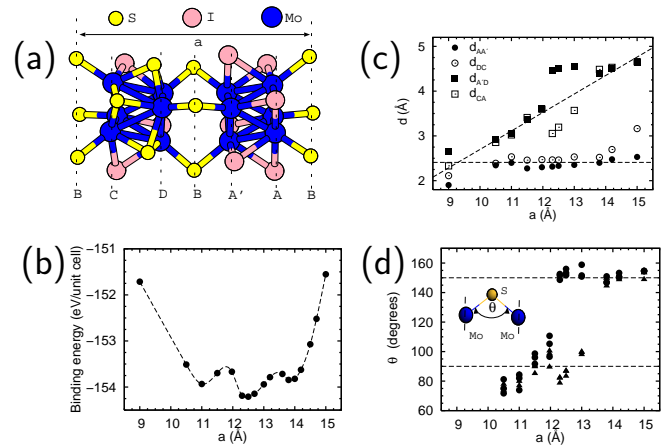}
\caption{(Color online) Structural properties of
Mo$_{12}$S$_{9}$I$_{9}$ nanowires. (a) Atomic arrangement within a
unit cell. Planes normal to the wire axis are denoted by A, A', B,
C, D. (b) Binding energy as a function of the lattice constant
$a$. (c) Optimized inter-plane distances as a function of $a$. (d)
Optimum values of the Mo-S-Mo bond angle $\Theta$ as a function of
$a$. Values in the two sulphur bridges per unit cell are
distinguished by different symbols. Dashed lines in (b-d) are
guides to the eye. \label{Fig1} }
\end{figure}


To gain insight into structural and electronic properties of these
unusual systems, we optimized the geometry of infinite
Mo$_{12}$S$_{9}$I$_{9}$ nanowires as well as selected structural
building blocks using density functional theory (DFT). We used the
Perdew-Zunger \cite{Perdew81} form of the exchange-correlation
functional in the local density approximation (LDA) to DFT, as
implemented in the \textsc{SIESTA} code
\cite{soler-tsmfaioms2002}. The behavior of valence electrons was
described by norm-conserving Troullier-Martins pseudopotentials
\cite{Troullier91} with partial core corrections in the
Kleinman-Bylander factorized form \cite{Kleinman82}. We used a
double-zeta basis, including initially unoccupied Mo$5p$ orbitals.
We arranged the nanowires on a tetragonal lattice, with one side
of the unit cell given by the lattice constant $a$. The other
sides of the unit cell were taken to be $19.7$~{\AA} long, about
twice the nanowire diameter, to limit inter-wire interaction. We
sampled the rather short Brillouin zone of these 1D structures by
6~k-points~\cite{MoSI-kpts}. The charge density and potentials
were determined on a real-space grid with a mesh cutoff energy of
$150$~Ry, which was sufficient to achieve a total energy
convergence of better than 2~meV/atom during the self-consistency
iterations.

Magnetic ordering in the nanostructures was investigated using the
Local Spin Density Approximation (LSDA) within the \textsc{SIESTA}
code \cite{soler-tsmfaioms2002}. Even though LSDA may be
considered less rigorous and dependable than LDA, we use it here
as a systematic way to estimate net magnetic moments and exchange
splitting in 1D nanowires and finite segments of these systems.

With 30 atoms per unit cell, corresponding to 84 degrees of
freedom, global structure optimization is a formidable task. The
initial experimental structure determination, based on
Atomic-Resolution Transmission Electron Microscopy and X-ray
diffraction data \cite{VNicolosi04ntconference}, provided us with
a useful starting point for determining the equilibrium structure.
In spite of the significant energy gain of 50~eV with respect to
the experimental structure during the initial structure
optimization, the calculated atomic arrangement within the
Mo$_{12}$S$_{9}$I$_{9}$ unit cell, depicted in Fig.~\ref{Fig1}(a),
was found to lie very close to the experimental structure.

We found it useful to assign atoms to planes normal to the wire
axis. We label these atomic layers A, A', B, C, and D in
Fig.~\ref{Fig1}(a). Layers A and A' are structurally identical in
terms of atomic arrangement, but rotated by 180$^\circ$ with
respect to each other. Mo octahedra, decorated by S and I atoms,
form the structural motif of the AA' and CD bilayers. As we
discuss later on, these bilayers can be thought of as rigid
blocks, connected by bridges formed of three S atoms, denoted as
layer B.

Our results for the binding energy of the system as a function of
the lattice constant $a$ are summarized in Fig.~\ref{Fig1}(b). For
lattice constants other than the experimental value
\cite{{VNicolosi04ntconference},{alat-exp}}
$a_{expt}=11.97$~{\AA}, the initial structures used in the
optimization were based on uniformly expanded or compressed
nanowires, subject to random distortions, or structures optimized
using semiempirical force fields. We considered a structure as
optimized when different starting geometries resulted in the same
structural arrangement.

The cohesive energy of the 30-atom unit cell, displayed in
Fig.~\ref{Fig1}(b), translates into a formidable average binding
energy exceeding 5~eV per atom. We found that inclusion of Mo$5p$
orbitals provided a net stabilization of the system,
but did not modify much the interatomic forces or equilibrium
geometries. As indicated by our LSDA calculations, additional
structure stabilization occurred due to magnetic ordering for
selected geometries.

Most intriguing in these results is the presence of multiple
structural minima. Since structures with optimum lattice constants
$a=11.0$~{\AA}, $12.3$~{\AA} and $13.8$~{\AA} are very close in
energy, a distribution of bonding configurations along the chain
axis will likely occur in order to maximize configurational
entropy, resulting in the loss of long-range order. Identifying
structures with very similar energies in a range of lattice
constants also implies the possibility to stretch the nanowire by
20\% at virtually no energy cost, similar to an accordion. The
remarkable ability of Mo$_{6}$S$_{x}$I$_{y}$ to expand easily by
nearly 30\% has been previously noted in Atomic Force Microscope
experiments on closely related systems \cite{AHassanien05pe}. This
uncommon flexibility of an inorganic nanowire may explain in
retrospect, why the observed lattice constant
\cite{{VNicolosi04ntconference},{alat-exp}} $a_{expt}=11.97$~{\AA}
may differ from the structural minima in Fig.~\ref{Fig1}(b).

Trying to understand the origin of the unusual elastic behavior of
Mo$_{12}$S$_{9}$I$_{9}$ nanowires, we first investigated the
inter-layer spacings as a function of the lattice constant.
Analysis of our data, presented in Fig.~\ref{Fig1}(c), suggests
that interatomic spacing within AA' and CD bilayers changes very
little with increasing lattice constant. The motif of the bilayers
is formed of six Mo atoms in near-octahedral arrangement, with
Mo-Mo bond lengths close to $2.6$~{\AA} within the layers and
$3.0$~{\AA} along the wire axis. In the following, we consider
these decorated octahedra as rather rigid building blocks of the
nanowires. We found the decoration order by I and S atoms to be
non-random, since an interchange of these atoms within the unit
cell raised the total energy by more than $0.5$~eV.

Most importantly, results in Fig.~\ref{Fig1}(c) suggested that
virtually all structural changes are accommodated by the sulfur
bridges, which form the A'-D and C-A connections. We found that
most changes occurred in the Mo-S-Mo bond angle $\Theta$, whereas
the Mo-S bond lengths remained nearly constant. Our results for
$\Theta$ as a function of the lattice constant $a$, shown in
Fig.~\ref{Fig1}(d), suggest that the expectation values of
$\Theta$ do not change uniformly, but rather group around
$90^\circ$ and $150^\circ$. In the following, we will call the
corresponding structural arrangements ``short'' and ``long''
bridges. Similar metastable structures have also been reported for
1D wires of group IV elements \cite{Senger05}. In the case of
isolated sulphur chains and S$_3$ molecules, we have also observed
a preference for similar bond angles as in Fig.~\ref{Fig1}(d). In
analogy with similar preferential bond angles found in different
allotropes of C and Si, we associate the short bridge with $sp^3$
and the long bridge with $sp^2$ hybridization.

In the case of Mo$_{12}$S$_{9}$I$_{9}$ nanowires, we find two such
sulphur bridges in each unit cell. Local minima in the cohesive
energy, plotted in Fig.~\ref{Fig1}(b), are found at the lattice
constants $a=11.0$~{\AA}, $12.3$~{\AA}, and $13.8$~{\AA}. The fact
that only three minima are found, and that $a=12.3$~{\AA} is close
to the average of the other two lattice constants, supports our
conclusion that there are only two types of S-mediated bonds
connecting the blocks, which are decoupled within the unit cell.
The three local minima in Fig.~\ref{Fig1}(b) thus correspond to a
short-short, short-long (or long-short), and a long-long bridge
configuration.

To verify that our optimum structures are not influenced by the
constraints of a periodic lattice with a fixed unit cell size, we
independently optimized the structure of the Mo-based building
blocks with the proper decoration by S and I atoms, with sulphur
bridge trimers attached at both sides. With the exception of the
bridge atoms, which changed their positions, we found the atomic
arrangement in Mo$_{6}$S$_{6}$I$_{6}$, modelling the AA' bilayer,
and Mo$_{6}$S$_{9}$I$_{3}$, modelling the CD bilayer, to lie close
to globally optimized nanowire structures, discussed in
Fig.~\ref{Fig1}.

Comparing the binding energies of the bilayer clusters,
$E_{coh}(AA')=-81.0$~eV and $E_{coh}(CD)=-89.2$~eV, we conclude
that the binding of sulphur atoms to the cluster is significantly
stronger than that of iodine atoms. We also found that the
isolated CD cluster acquires a net magnetic moment
${\mu}(CD)=1.00~\mu_0$. As we discuss in the following, not only
finite nanostructures with dangling bonds, but also infinite
nanowires may develop a net magnetic moment.


\begin{figure}[tbp]
\includegraphics[width=0.81\columnwidth]{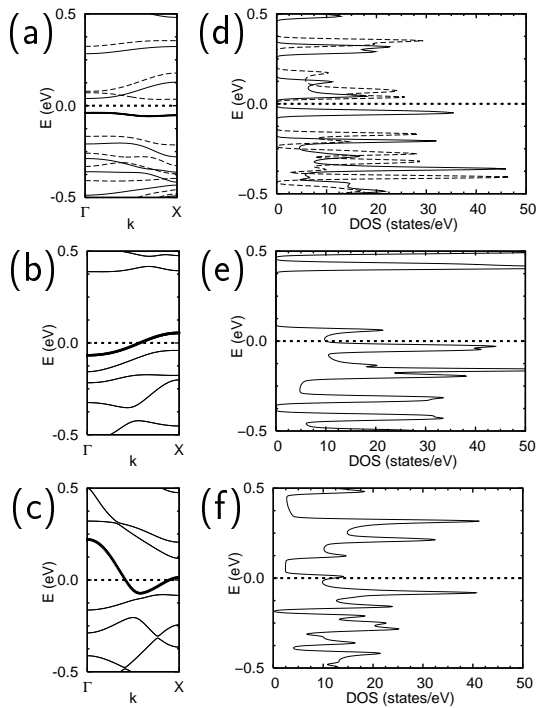}
\caption{ Electronic band structures $E(k)$ of
Mo$_{12}$S$_{9}$I$_{9}$ nanowires with lattice constants (a)
$a=11.0$~{\AA}, (b) $a=12.3$~{\AA}, and (c) $a=13.8$~{\AA}. Solid
lines represent majority and dashed lines minority spin bands.
Bands crossing the Fermi level are emphasized by heavy lines.
Densities of states of nanowires with (d) $a=11.0$~{\AA}, (e)
$a=12.3$~{\AA}, and (f) $a=13.8$~{\AA}, convoluted with $0.01$~eV
wide Gaussians. The Fermi level lies at $E_F=0$. \label{Fig2} }
\end{figure}


We found the infinite Mo$_{12}$S$_{9}$I$_{9}$ nanowires to
generally behave as conductors in the lattice constant range
$9$~{\AA}$<a<15$~{\AA}, with the exception of the metastable
structure at $a{\approx}11$~{\AA} discussed below. In
Fig.~\ref{Fig2} we present results for the electronic structure at
$a=11.0$~{\AA}, $12.3$~{\AA}, and $13.8$~{\AA}, corresponding to
the equilibrium structures indicated in Fig.~\ref{Fig1}(b). The
interactions along the nanowire, causing band dispersion, depend
crucially on the hybridization near the sulphur bridges. As
discussed earlier, changing the lattice constant does not affect
interatomic distances, but rather modifies the Mo-S-Mo bond angle
$\Theta$, thus changing the hybridization of directional orbitals.
An increase in band dispersion, caused by increased hybridization,
may be counter-acted by band repulsion in complex systems.

The interplay between lattice constant and band structure is
depicted in Figs.~\ref{Fig2}(a-c). Our results in
Figs.~\ref{Fig2}(b-c) suggest that the metallic character of the
wires derives from a rather dispersionless band, which crosses the
Fermi level. As shown in Fig.~\ref{Fig2}(a), this band becomes
very flat and even changes its slope at $a{\approx}11.0$~{\AA} due
to the changed hybridization associated with Mo-S-Mo bond bending.
Consequently, the density of states develops a peak at $E_F$,
which subsequently splits due to a magnetic instability, as shown
in Fig.~\ref{Fig2}(d). This splitting opens up a fundamental gap
of 73~meV, accompanied by a magnetic moment of $1.00~\mu_0$ per
unit cell. At other lattice constants, the dispersion of this
partly filled band increases, as seen in Figs.~\ref{Fig2}(b) and
(c). Consequently, as seen in Figs.~\ref{Fig2}(e) and (f), the
density of states at the Fermi level is lower, suppressing the
magnetic instability.

\begin{figure}[tbp]
\includegraphics[width=0.95\columnwidth]{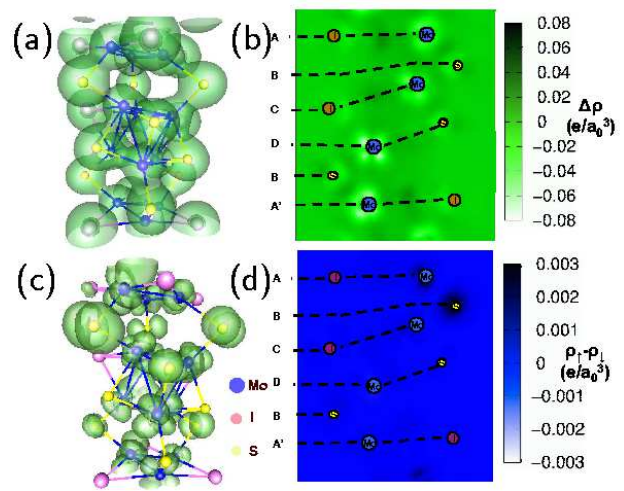}
\caption{(Color online) Contour plots depicting the charge
distribution in a Mo$_{12}$S$_{9}$I$_{9}$ nanowire at
$a=11.0$~{\AA}. (a) Total charge density contour at
${\rho}=0.05$~el/a$_{0}$$^3$. (b) Difference charge density
$\Delta{\rho}(\textbf{r})$, indicating regions of charge depletion
and excess with respect to the superposition of isolated atoms.
(c) ${\rho}=1.0{\times}10^{-3}$~el/a$_{0}$$^3$ charge density
contour associated with states close to the Fermi level,
$E_F-0.05$~eV$<E<E_F+0.05$~eV. (d) Difference spin density
$\rho_\uparrow(\textbf{r})-\rho_\downarrow(\textbf{r})$, showing
regions of excess majority spin $\uparrow$. Contour plots (b) and
(d) are depicted in a plane, which contains the wire axis.
\label{Fig3} }
\end{figure}


A more detailed discussion of the electronic structure in the
system with a net magnetic moment, at $a=11.0$~{\AA}, is presented
in Fig.~\ref{Fig3}. The charge distribution in the
Mo$_{12}$S$_{9}$I$_{9}$ nanowire, depicted in Fig.~\ref{Fig3}(a),
is very similar to the superposition of atomic charge densities.
This is better seen in Fig.~\ref{Fig3}(b), which displays the
charge redistribution within the system with respect to isolated
atoms, given by $\Delta{\rho}(\textbf{r})=\rho(\textbf{r})
-\Sigma\rho_{\text{\textrm{atom}}}(\textbf{r})$. As expected based
on electronegativity differences, we observe a small amount of
charge, stemming mostly from Mo$4d$ orbitals, to be transferred to
the region within the Mo octahedra, in particular to Mo-Mo bonds
and sulphur atoms. The decorating iodine atoms do not experience a
change in their charge distribution due to their presence in the
nanowire structure.
More interesting than information about all populated levels is
the spatial distribution of states close to the Fermi level, which
form the valence and conduction band. Our results, shown in
Fig.~\ref{Fig3}(c), suggest that these states are rather
delocalized.
Together with the band structure results of Fig.~\ref{Fig2}, we
conclude that the nanowires considered here should be conducting
at room temperature, in agreement with experimental observations
\cite{{VNicolosi05cpl},{VNicolosi05jpcb}}.

Spatially resolved information about the magnetic structure of the
nanowire at $a=11.0$~{\AA} is presented in Fig.~\ref{Fig3}(d).
Associating up-spin with the majority and down-spin with the
minority states, the plotted quantity
$\rho_\uparrow(\textbf{r})-\rho_\downarrow(\textbf{r})$ allows us
to identify spatial regions with a net magnetic moment. Based on
the results in Fig.~\ref{Fig3}(d) and a set of 3D spin density
contours, not shown here, we find the majority spin to be
localized mostly on the bridging sulphur atoms, in their
non-bonding $2p$ orbitals normal to the wire axis, and to a
smaller degree on adjacent Mo atoms.

Since magnetism occurs in selected structural isomers of the
infinite Mo$_{12}$S$_{9}$I$_{9}$ nanowires as well as in isolated
building blocks, we expect the possibility of modifying the
magnetic structure by locally stretching or expanding the
nanowires. Since such structural changes may be induced locally at
negligible energy cost, various magnetic patterns could possibly
be obtained by mechanical manipulation of Mo$_{12}$S$_{9}$I$_{9}$
nanowires, possibly using an Atomic Force Microscope. Since
magnetic ordering will likely affect spin transport in these
systems, Mo$_{12}$S$_{9}$I$_{9}$ nanowires may be viewed as spin
valves, which could change their state by applying local stress.

We note that the overall metallic behavior of
Mo$_{12}$S$_{9}$I$_{9}$ nanowires is rather unusual among
chalcogenide structures. Significant interest has been devoted in
the past to the ability of layered semiconductors, such as
MoS$_2$, to form fullerene- and nanotube-like structures
\cite{{Feldman95},{Seifert00},{Remskar01sci}}. Similar to their
bulk counterparts, MoS$_2$ nanotubes were found to be
semiconducting, unless intercalated with iodine to enhance
conductance \cite{{Charlier03prb},{AZimina04nl}}. MoS$_2$-based
fullerenes and nanotubes, often containing iodine, were noted in
particular for their favorable tribological properties
\cite{{Feldman95},{Remskar01sci},{Pottuz05trl}}. At other
stoichiometries, such as Mo$_{6}$S$_{3}$I$_{6}$ or
Mo$_{6}$S$_{4}$I$_{5}$, self-assembled nanowires have been
reported to form with either semiconducting or semi-metallic
behavior \cite{{DVrbanic04nt},{AMeden05nt}}. In the particular
Mo$_{12}$S$_{9}$I$_{9}$ stoichiometry studied here, the rich
behavior of molybdenum chalcogenides is complemented by new
properties, such as high structural flexibility,
metal-semiconductor transition, and magnetism.
We propose to use these systems as unique building blocks in
hierarchically assembled nanostructured electronic circuits.


We acknowledge discussions and collaboration with Valeria Nicolosi
and Jiping Li during the initial stages of this work. Partial
funding was provided by the NSF NIRT grants ECS-0103587,
ECS-0506309, and NSF NSEC grant EEC-425826.


\begin{thebibliography}{22}
\bibitem[{MoS()}]{MoSI-kpts}
\bibinfo{note}{Details of the electronic structure were obtained using a finer
  mesh of $16$~k-points.}
\bibitem[{ala()}]{alat-exp}
\bibinfo{note}{Since the X-ray diffraction peaks in the direction of
  Mo$_{12}$S$_{9}$I$_{9}$ nanowires appear somewhat diffuse, we believe that
  the long-range order, including the lattice constant, is not as well defined
  as the short range oder.}
\end{thebibliography}

\end{document}